# LISA-2020: An Intermediate Scale Space Gravitational Wave Observatory for This Decade

S. Buchman[1], J.A. Lipa[1], R.L. Byer[1], D. DeBra[1], K. Balakrishnan[1], G. Dufresne Cutler[1],
A. Al-Fauwaz[1,2], E. Hultgren[1], A.K. Al-Jadaan[1,2], S. Saraf[1], S. Tan[1], S. Al-Thubiti[1,2], A. Zoellner[1]

[1] W.W. Hansen Experimental Physics Lab, Stanford University, Stanford CA 94305, USA
[2] King Abdulaziz City for Science and Technology, Riyadh 11442, Saudi Arabia

Over the last three decades, an exceptionally good science case has been made for pursuing gravitational wave (GW) astronomy. This has engendered a worldwide effort to detect the extremely weak signals generated by expected sources. With the next round of upgrades the ground based instruments are likely to make the first detections of the sources, and a new era of astronomy will begin, possibly as early as 2017. Inconveniently, due to seismic noise and baseline length issues, the low frequency (<10Hz) part of the spectrum, where the most interesting events are expected, will not be accessible. The space-based detector, LISA[1], was conceived to fill this gap extending the observational capability to about $10^{-4}$ Hz. Due to mission cost growth and severe budget constraints, a flight prior to 2030 now seems very unlikely. This paper examines the case for a scaled down mission that is comparable in cost and duration to medium scale astrophysics missions such as the 1978 ($630M) Einstein (HEAO 2) x-ray Observatory[2], the 1989 ($680M) COBE Cosmic Background Explorer[3], and the 1999 ($420M) FUSE Far Ultraviolet Spectroscopic Explorer[4]. We find that a mission of this class is possible if the measurement requirements are somewhat relaxed and a baseline smaller than LISA is used. It appears that such a mission could be launched by 2020 using a conventional program development plan, possibly including international collaboration. It would enable the timely development of this game-changing field of astrophysics, complementing the expected ground results with observations of massive black hole collisions. It would also serve as a stepping stone to LISA, greatly reducing the risk profile of that mission.

**The case for GW Astronomy**

Our present understanding of the Universe rests primarily on imaging in the electromagnetic spectrum, spanning radio to gamma ray astronomy. Visible matter generating light represents only about 0.4% of the total mass of the Universe, with dark energy 73%, dark matter 23% and intergalactic gas 3.6% constituting the rest. Of the four fundamental forces, two are long range and allow remote observation: electromagnetism and gravitation. Electromagnetic radiation is generated by electrons in 0.4% of matter, is readily absorbed and/or scattered and does not interact significantly with dark matter and dark energy - the presumed constituents of 96% of the Universe, Figure 1 (left).

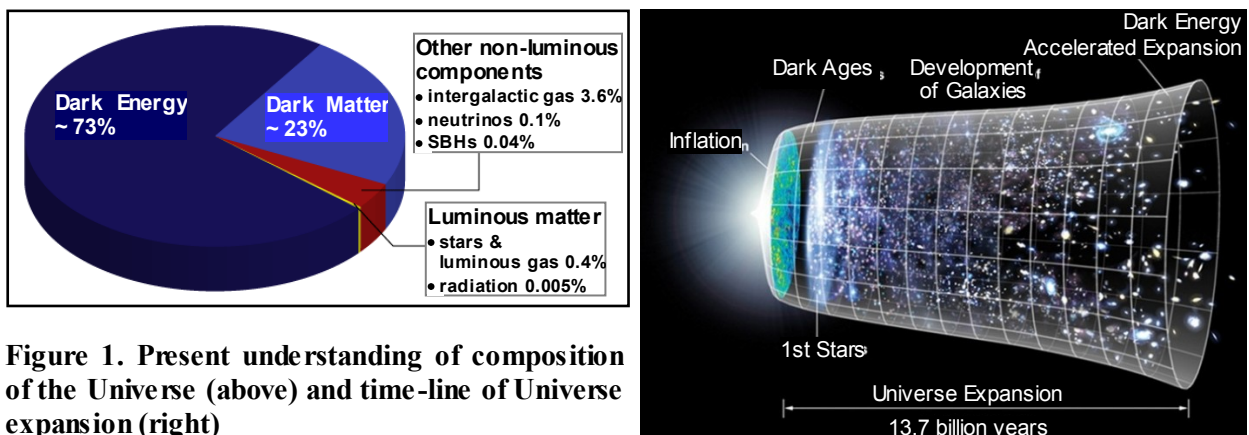

Figure 1. Present understanding of composition of the Universe (above) and time-line of Universe expansion (right)

Although extraordinarily productive from the dawn of humanity, the information carried by light is limited in its capability to inform us about a Universe that gets 'darker' and more complex as scientific discovery progresses. While the weakest of all forces, gravitation, maps the curvature of space-time, interacts with all matter and is the driving force that shapes the Universe. GWs permeate space-time at the speed of light and are not attenuated in their propagation. The detection of GWs will open a new door to understanding the origin, evolution, and future of the Universe. In particular, ground based gravitational wave antennas like LIGO[5] are expected to observe GWs beginning in 2017 but be limited to higher frequencies and a limited detection range of about 100 Mpc. On the other hand, space-based gravitational antennas, with a lower frequency detection capability in the $\sim 10^{-4}$ Hz to 1 Hz range and with improved sensitivity, will probe gravitational wave interactions of Massive black hole (MBH) binaries over the entire space-time range Universe. GW antennas in space thus represent a new and fundamental tool for probing the frontiers in our understanding of the Universe.

Moreover, in the last few decades physics and astrophysics have been raising an increasing number of questions about Einstein's theory of General Relativity (GR) that remains, one hundred years after its formulation, the dominant accepted scientific description of gravitation and is fundamental to the interpretation of electromagnetic astrophysics. Major GR 'difficulties' include:

a) GR is experimentally verified to about $10^{-5}$; a relatively low precision compared to other forces and fundamental constants[6].

b) GR is not yet unified with the SM and/or GUT: GR is not integrable in the Standard Model (SM) or even with the Grand Unified Theories (GUTs)[7] of the electromagnetic, weak and strong forces; note that none of the GUTs are generally accepted.

c) The original GR of the stationary Universe, $R_{\mu\nu} - \frac{1}{2}Rg_{\mu\nu} + \Lambda g_{\mu\nu} = T_{\mu\nu}$ required a cosmological constant $\Lambda$, that was later removed by Einstein, to accommodate the uniformly expanding Universe, only to be resurrected as one explanation of the recently discovered acceleration in the expansion of the Universe, see Figure 1 (right).

The 2017 upgrades to the ground based LIGO interferometers will improve sensitivity by 10, the event rate by 1000 and the detection range to about 0.1 Gpc. The expectation of the community is that LIGO is then likely to detect GW sources. At present, the first space-based gravitational wave antenna, in the most optimistic scenario, will not be functional before 2030 at a cost in excess of $2 billion. The proposed schedule and cost make such a program unrealistic as follow-on to the likely progress in ground-based observations and unattractive to the contemporary generation of young scientists.

Here we propose a space-based gravitational wave antenna called LISA 2020 with a launch date near 2020 and a cost in the $500 million range. The LISA 2020 gravitational wave antenna is based on three principles: 1) operate at a sensitivity to achieve most of the science goals of the LISA mission; 2) keep costs down by developing and testing critical technologies in parallel, utilizing small satellites; and 3) engage the international community in the effort to develop the instrument and mission in a community-wide science program. We show that LISA 2020 is implementable at a cost of about $0.5 billion and can be flown by 2020 using a combination of parallel developments with selectively reduced sensitivity and significantly reduced complexity.

**The LISA 2020 concept**

Modern GW detectors are based on the measurement of the modulation of space-time caused by the passing of a GW between two or more 'free floating' test masses (TM). In 1971 R. Weiss[5] first promoted the concept of laser interferometry as the best method to achieve the precision required for the detection of GW. Interferometers operating at the quantum noise limit are now the instrument of choice for ground-based detectors. The first space-based detector was proposed in 1980 similarly based on laser interferometry. Conceptually, a GW detector is a modified Michelson interferometer consisting of two TM and a ruler based on light. The difficulty arises due to the 'weakness' of the gravitational interaction and the 'stiffness' of space. GW amplitudes, known as 'strain amplitude $h$' and defined by $h \equiv dl/l$, are typically expected to be $h \approx 10^{-20}$ for sources detectable by ground-based GW detectors. Measuring the strain of space-time to this precision is equivalent to measuring the displacement of an atom at the distance of the Sun.

GW detectors in space benefit from very long baselines available and the absence of seismic-type noise generated on the ground. A 5 million kilometer baseline distance between spacecraft allows the detection of GWs to frequencies of $10^{-4}$ Hz. Figure 2 shows the strain sensitivities of LIGO and LISA, the ground and space-based interferometers. The 4 km arm length of LIGO, limited by the curvature of the Earth, puts the detection band in the 10 to 1000 Hz range. The 3 million km arm length of the original LISA interferometer puts the detection band in the $10^{-4}$ to 1 Hz range. Figure 2 also shows the sources that generate the GWs for both LIGO and LISA.

Three main factors contribute to the performance and cost of a space detector: 1) the orbit of the three-spacecraft constellation, 2) the design of the drag-free test masses (TM) contained within the satellites, and 3) the laser interferometer measurement system operating over millions of km to picometer precision. LISA, the benchmark space detector, has been studied and developed since 1993, and is budgeted at $2.1 billion dollars[8]. In 2011 a number of alternatives were investigated by NASA[8]. As part of that study, an international collaboration proposed a mission called LAGRANGE (Laser Gravitational-wave ANtenna in GEocentric orbit)[9] with an orbit at the L3, L4, and L5 Lagrange points of the Earth-Moon system, and with arms having **0.67 million km lengths**. The advantages of the geocentric orbit coupled with a single spherical TM per spacecraft with an acceleration noise of less than **$3 \times 10^{-15}$ m·s$^{-2}$·Hz$^{-½}$** and a measurement precision of **8 pm·Hz$^{-½}$** led to strain sensitivity of $10^{-19}$ at 0.01 Hz.

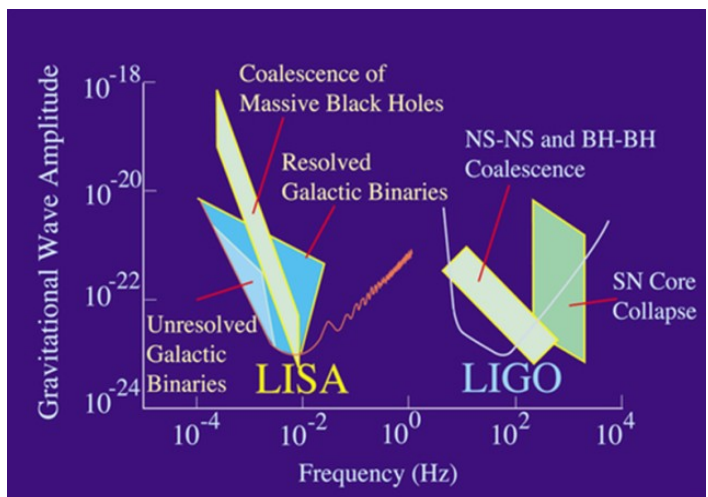

**Figure 2** The LISA and LIGO sensitivity to GW strain vs frequency and the GW sources at a signal to noise for one year of signal averaging

A geocentric orbit (0.67 Gm to 1.0 Gm arms) presents decisive advantages over a heliocentric one by reducing the launch weight by half and increasing the telemetry and command bandwidth by a more than two orders of magnitude. Augmented requirements for thermal control and Doppler shift compensation are well within the present technology capabilities of active-control multi-layer insulation and phase-meters. Single spherical TM's per spacecraft have a long flight heritage and further decrease the complexity and weight of the experiment. Note that alternate TM and interferometry designs can replace the proposed ones if these systems have reached high Technological Readiness Level (TRL) and are cost effective. LAGRANGE has been estimated to cost between $0.65 billion and $1 billion, in excess of the $0.5 billion sought by NASA. Note however that most of LISA science remains available at a projected cost reduction of more than 50%.

Presently, subject to numerous caveats, ESA has considered flying a two-arm LISA-like mission (eLISA)[10] at a cost of 1.5 billion dollars[11], but '*not before*' 2028, while NASA has similar plans for the 2030's. We here propose LISA 2020; a name that combines the widely recognized LISA trademark with a launch time that is both a challenge and a promise to the community. LISA 2020 is based on the LAGRAGE general concepts, but reduces the sensitivity requirements of drag-free performance and interferometry by a **factor of 30**. The TM drag-free requirement is relaxed from **$3\times10^{-15}$ m·s$^{-2}$·Hz$^{-½}$** to **~$10^{-13}$ m·s$^{-2}$·Hz$^{-½}$** and the interferometer performance from **8 pm· Hz$^{-½}$** to **~250 pm·Hz$^{-½}$**. This level of technology is well established[12,13] and prototypes can be demonstrated on mini satellites. Relaxation of requirements results in the relaxation of environmental, optical, mechanical control and electronics requirements across the board including thermal stability, magnetic susceptibility, pressure at TM, mass distribution uniformity and stability, EMI, optical path length control and mechanical stability.

The program would start with a 3-4 year campaign of four to six critical technology demonstrations in parallel on small satellites. A core team would design and start building the *well known satellite and science instrument* components: satellites, telescopes and standard electronics. A review would establish the TRL of the critical technologies and the program would proceed as appropriate with building the instruments over 3-4 more years following the establishment of appropriate TRLs via small satellite missions. The relaxed requirements coupled with the further reduction in weight from LAGRANGE (from 2,000 kg total to about 1,000 -1,500 kg) will bring LISA 2020 into the '*affordable*' $0.5 billion range. Optimally the program would be directed by a team of academic scientists, on the model of LIGO or Fermi.

**LISA 2020 science**

Figure 3 (left), adapted from ref[14] shows the estimated LISA 2020 strain sensitivity (dotted black curve) in units of Hz$^{-1/2}$ and a cartoon of the LISA 2020 constellation (right). Colored curves and points represent the amplitude spectral density of gravitational wave strain versus frequency for the various known sources within this bandwidth[14]. About half dozen known galactic binaries will ensure calibration of the instrument. Compared to LISA, the MBHB sources are fully detectable, albeit at a lower though still large signal to noise ratio. Extreme Mass Ratio Inspirals ( EMRIs), are most likely not detectable according to present cosmological models. All sources above the sensitivity curve are detectable by LISA 2020. Detection rates for resolved galactic binaries are about 10% that of LISA, though still well in excess of $10^3$ per year. Total rates of MBHB detected mergers are reduced to 20% - 40% of LISA expectations, resulting in a likely 20 to 40 detections per year (including 1 to 5 with $z>10$)[15].

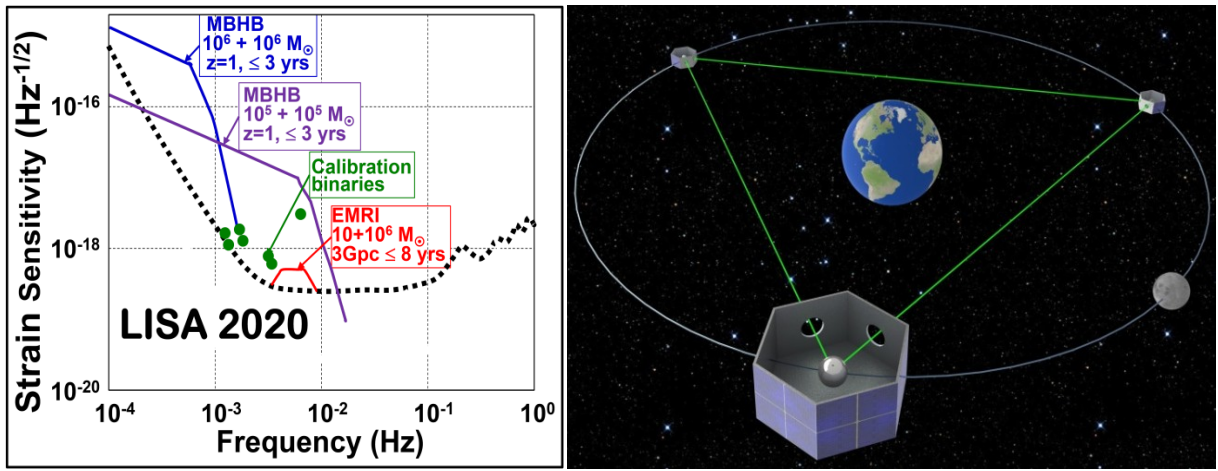

**Figure 3. LISA 2020 strain sensitivity per Hz$^{½}$ (left) and orbital configuration (right)**

Note that the GW frequency from MBHB and EMRI sources increases during observation. Sensitivity normal to the ecliptic plane is less than that of LISA due to the reduced out-of-plane motion of the observatory. However, higher gravitational-wave harmonics due to the shorter baseline of the LISA 2020 constellation provide a significant improvement in the position determination of MBHBs. Locating spinning black holes in a MBHB is much more accurate than would be expected from the modulation produced by the LISA precessing plane alone[16].

**Conclusions**

LISA 2020 will achieve the most important science goals of LISA listed in the 2010 astrophysics decadal survey, "New Worlds, New Horizons" [17]:

(i) Measurements of black hole mass and spin from MBHB will be important for understanding the significance of mergers in the building of galaxies.

(ii) LISA 2020 is unlikely to detect the signals from stellar-mass compact stellar remnants as they orbit and fall into massive black holes (EMRI) which would provide exquisitely precise tests of Einstein's theory of gravity. However, an equally powerful test will be provided by the mergers of MBHB by comparing actual GW forms to the highly detailed numerical simulations performed by modern general relativistic hydrodynamics codes with dynamical space-time evolution[18].

(iii) Potential for discovery of waves from unanticipated or exotic sources, such as backgrounds produced during the earliest moments of the universe, dark energy signals or cusps associated with cosmic strings. Consequently, LISA 2020 will, with relaxed requirements, deliver excellent science in a timely fashion and at a cost affordable by an international consortium.

Our knowledge of the Universe is critically dependent on the understanding of gravitation by means of GWs; our only available tool for direct observation of gravitation. A combination of high and low frequency GW observations, from ground-based detectors and missions such as LISA 2020, is highly desirable to achieve early in the next decade the next key breakthroughs in our understanding of the new and dark Universe hinted at by electromagnetic wave astronomy. LISA 2020 builds on the ground based LIGO observatory and adds significant new sources and science to GW astronomy in a timely way. Note that the timing of this program will open the opportunity to take advantage of current capabilities and to involve a number of current participants in the LIGO project, including 400 data analysis scientists and more than 50 optical physical scientists and engineers. LISA 2020 would be a bridge to higher sensitivity GW antennas, but in the near future and at a cost comparable with other mid-sized Astrophysics missions.

**References**

[1] LISA – *Pre Phase A Report*, December 1995, MPQ 208 (1996)
[2] R. Giacconi et al., *Astrophysical Journal*, **230**, 540 (1979)
[3] J.C. Mather, et al., Adv. Space Res., **11**(2), 181 (1991)
[4] D.A. Content, et al., *Instrumentation in Astronomy VII* **1235**, 943 (1990)
[5] R. Weiss, *Electromagnetically Coupled Broadband Gravitational Antenna* - Quarterly Progress Report of RLE, MIT **105**, 54 (1972)
[6] C. M. Will, *Living Rev. Relativity*, **9**, (2006)
[7] H. Georgi, S.L. Glashow, *Phys. Rev. Let.* **32**, 438 (1974)
[8] National Aeronautics and Space Administration, Gravitational-Wave Mission Concept Study Final Report August 9, 2012
[9] J W Conklin, et al. LAser *GRavitational-wave ANtenna at GEo-lunar Lagrange points* (2011)
[10] Gravitational Wave Astronomy in Space eLISA/NGO
[11] European Space Agency, NGO Revealing a hidden Universe: opening a new chapter of discovery, Assessment Study Report, ESA/SRE(2011)19, December 2011
[12] Adam J. Mullavey et al, *Arm-length stabilisation for interferometric gravitational-wave detectors using frequency-doubled auxiliary lasers*, Optics Express, Vol. **20**, Issue 1, pp. 81-89 (2012)
[13] F Antonucci et *al, From laboratory experiments to LISA Pathfinder: achieving LISA geodesic motion*, arXiv:1012.5968v3 [gr-qc]

14. R. Stebbins et al. Laser Interferometer Space Antenna (LISA) A Response to the Astro2010 RFI for the Particle Astrophysics and Gravitation Panel, white paper. 2009.
15. I. Thorpe, J. Livas, *NASA's Gravitational---Wave Mission Concept Study For the GW Study Team Physics of the Cosmos Program Analysis,* Group Meeting Washington DC, August 14th, (2012)
16. C. L. Wainwright and T. A. Moore, *Phys. Rev. D*, **79**(2):024022, (2009)
17. Committee for a Decadal Survey of Astronomy and Astrophysics; NRC New Worlds, New Horizons in Astronomy and Astrophysics, The National Academies Press, Washington, D.C. (2010)
18. B. Zink, et al, Phys. Rev. Lett., 96, 161101, (2006)